# Sound propagation in a solid through a screen of cylindrical scatterers

Y.C. Angel

*University of Lyon, Lyon, F-69003, France; University Lyon 1, Lyon, F-69003, France; INSERM, U556, Therapeutic Applications of Ultrasound Laboratory, Lyon, F-69424, France.*

C. Aristégui, M. Caleap

*Université de Bordeaux, CNRS, UMR 5469, Laboratoire de Mécanique Physique, Talence, F-33405, France.*

**Abstract**

The propagation of SH waves in a solid containing a screen of line-like scatterers is investigated. When the scatterers are uniformly distributed, the amplitudes of the coherent waves inside and outside the screen are evaluated in closed form. In the analysis, multiple scattering effects are taken into account within the context of a first-order approximation. A Global Closure Assumption is proposed, which yields an effective wavenumber identical to that of Waterman and Truell. The scatterers can be fibers of circular or elliptical cross-sections; they can also be two-dimensional cracks with slit-like or elliptical cross-sections. Specific analytical and numerical results are presented for flat cracks and empty cavities of circular cross-sections. In those two cases, figures are presented to illustrate the variations of the reflection and transmission coefficients as functions of frequency and of scatterer concentration. The crack and cavity results, respectively, are compared with those of earlier works.

## 1. Introduction

Waves propagating in biological, or structural materials, containing fibers are subjected to multiple scattering when the wavelengths of the probing signals are sufficiently short relative to a characteristic cross-length of the fibers. With high fiber concentrations, diffusive effects may also be observed (Ishimaru 1997). Here, we are interested in ultrasound measurements in materials with low fiber concentrations, where multiple-scattering effects can be significant and must be taken into account, if correct interpretations are to be inferred from reflected or transmitted signals.

Experimental and analytical difficulties associated with the phenomenon of multiple scattering have been dealt with for many years. Yet, the task is complicated, and the phenomenon is still being investigated, both experimentally and analytically (Challis *et al.* 2005; Derode *et al.* 2006; Kim 2003, 2004; Le Bas *et al.* 2005; Linton & Martin 2005, 2006; Wang & Gan 2002; Wilson *et al.* 2005).

Here we look at analytical aspects of the problem. We derive concise and simple expressions for reflection and transmission (back-scattering and forward-scattering). These expressions will provide a tool to test, in the laboratory, the validity of the theory.

The present work applies to the case of SH (antiplane) waves in a linearly-elastic solid. Scatterers are solid fibers or cavities, of circular or elliptical cross-sections. They can also be two-dimensional cracks with slit-like or elliptical cross-sections. Solid fibers can be coated or layered. Propagation is perpendicular to the cylinder axes, and particle displacements are parallel to those axes. Thus, the displacements are solutions of scalar wave equations, and we can talk of "scalar waves."

Our analytical approach follows that of Foldy-Twersky and of Waterman-Truell (Ishimaru 1997). Their equations are reexamined in detail. In particular, their closure assumption concerning the field scattered by a fixed scatterer in the presence of all other scatterers is reformulated and labeled "Global Closure Assumption" in Section 6. This allows us to state, in mathematical terms, the approximation that is made to arrive at the solution of the problem. In the end, we obtain an expression for the effective wavenumber that is identical to that of Waterman and Truell (1961).

Our work is an extension of theirs. Indeed, we determine closed-form expressions for the reflection and transmission coefficients – which they had not obtained. These expressions depend on scatterer concentration and size, frequency, the properties of the solid matrix, the thickness of the scattering region, as well as on the forward and backward scattering properties of a *single* scatterer.

Discussions available in the literature seem to agree that closure assumptions such as that of Waterman and Truell or such as our Global Closure Assumption yield predictions that are in good agreement with experimental measurements – but only in the limit of low scatterer concentration. However, to the best of our knowledge, this observation has always been made on limited sets of experimental data. There exist no extensive data collections of ultrasonic measurements for a wide variety of materials containing scatterers.

This work on SH waves follows an earlier work where P waves are considered (Angel & Aristégui 2005). Because of more immediate potential applications in (fluid-like) biological tissue, we started with P waves, which are also called "vector waves". In Angel & Aristégui (2005), scalar displacement potentials are used in the formulation of the problem. Here, for SH waves, potentials are not needed. Thus, one cannot simply infer from Angel & Aristégui (2005) the results of this paper, although there are some similarities between the two works. For this reason, we have included in Section 9 a comparative discussion of the two methods of solution. In that section, we point out that the amplitude formulae for the SH and P problems, respectively, have been made to look identical. To arrive at those expressions, however, the respective $Q$ factors must be defined differently (see (9.33) and (9.34)). In Angel & Aristégui (2005), the theory only is developed. No numerical examples are presented.

We present in Section 10 numerical results corresponding to a distribution of flat cracks and a distribution of cylindrical cavities, respectively. The results are compared with those of earlier works, where scatterers are not replaced with equivalent line scatterers. We also show numerical predictions for the reflection from a half-space containing a distribution either of flat cracks or of cylindrical cavities.

## 2. Screen of scatterers

We consider an SH wave that encounters cylindrical scatterers, as shown in Fig. 1. The scatterers, of identical geometry and elastic properties, are uniformly and randomly distributed in a layer of thickness $2h$, where their number concentration per unit area in the $(y_3, y_1)$ plane is $n_0$. Interactions with the scatterers cause the incident wave to be reflected and transmitted on either side of the layer.

For a given incident wave and a given number concentration $n_0$, there exists an infinite number of scatterer configurations and an infinite number of corresponding wave motions. The configuration average of these wave motions is a weighted sum, where the weight represents the probability of finding a given configuration among all possible configurations.

The scatterers have cylindrical geometry, with constant cross sections in the $(y_3, y_1)$ plane of Fig. 1 and cylinder axes in the $y_2$ direction. We assume further that the cross sections are symmetrical with respect to axes parallel to the $y_3$ axis.

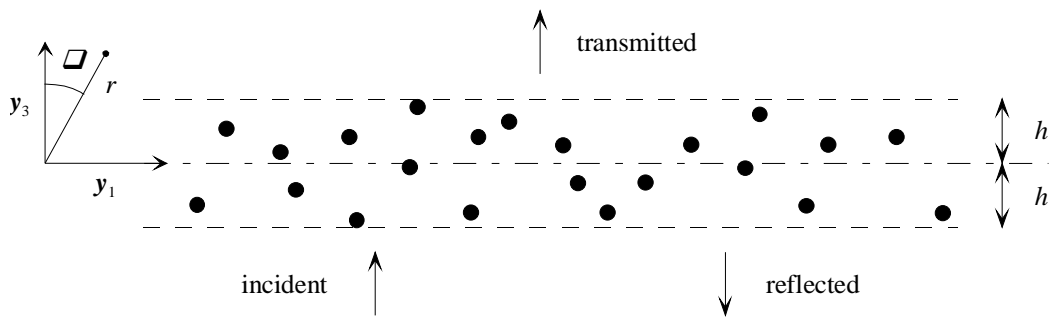

Figure 1 : Layer of scatterers and system of acoustic waves. Particle displacements are in the $y_2$ direction.

It follows from the preceding assumptions on the distribution and geometry of the scatterers that the wave system generated by the SH wave in Fig. 1 is two-dimensional. The solid particles move in the $y_2$ direction, with $u_2$ the displacement measured from the equilibrium configuration. The stress components in the context of small elastic deformations are given by

$$s_{21} = \mu \frac{\partial u_2}{\partial y_1}, \quad s_{23} = \mu \frac{\partial u_2}{\partial y_3}, \tag{2.1}$$

where $\mu = \rho_0 c_0^2$ is the shear modulus of the solid. We denote by $\rho_0$ the mass density and by $c_0$ the speed of transverse waves in the solid.

If the time dependence is of the form $\exp(-i\omega t)$ and the regime is continuous, the three waves of Fig. 1 are represented, respectively, by the three displacements

$$u_2^i = u_0 e^{iky_3}, \quad u_2^r = u_0 R e^{-iky_3}, \quad u_2^t = u_0 T e^{iky_3}, \tag{2.2}$$

where $k = \omega/c_0$ is the wavenumber, $u_0$ is an amplitude factor, $R$ is the reflection coefficient, $T$ is the transmission coefficient, and the factor $\exp(-i\omega t)$ has been omitted. The waves (2.2) propagate in the solid outside the layer. Inside the layer, where the scatterers are distributed, there are two waves traveling, respectively, in the forward and backward directions (as we shall see later).

Both $R$ and $T$ in (2.2) are *a priori* frequency dependent. They represent, respectively, averages on disorder of the amplitudes of the backscattered and transmitted waves generated by a continuous harmonic wave. These amplitudes depend also on the number concentration $n_0$, the size and material properties of the scatterers, the density and speed of sound in the surrounding solid, and on the layer thickness $2h$. Another parameter is the angle of incidence, but in this work we focus attention on normal incidence.

The scattering properties of a single scatterer are described by a set of coefficients $C_n$. Calling $u_2$ the scattered displacement in the solid, one has

$$u_2(r,q) = \sum_{n=0}^{+\infty} e_n A_n C_n H_n^{(1)}(kr)\cos(nq). \tag{2.3}$$

This equation is written in terms of the cylindrical Hankel function $H_n^{(1)}$. The Neumann factor is defined by $e_0 = 1$ and $e_n = 2$ for $n > 0$. The coefficients $A_n$ characterize the displacement of the exciting wave on the scatterer. In equation (2.3), the variables $(r, q)$ are local cylindrical coordinates centered on the axis of the scatterer. Let $\boldsymbol{x}$ denote the position of this axis in the coordinate system of Fig. 1. Then, the distance $r$ is defined by $r^2 = (y_3 - x_3)^2 + (y_1 - x_1)^2$. The angle $q$ is such that $\cos q = (y_3 - x_3)/r$. Corresponding to the incident wave of (2.2), one finds that the coefficients $A_n$ are

$$A_n = u_0 i^n e^{ikx_3}. \tag{2.4}$$

These coefficients are those of the incident displacement in the expression

$$u_2^i(r,q) = u_0 e^{ikx_3} e^{ikr\cos q} = \sum_{n=0}^{+\infty} e_n A_n J_n(kr)\cos(nq), \tag{2.5}$$

where $J_n$ denotes the cylindrical Bessel function.

The rule (2.3) and (2.5) says that the n[th] mode of amplitude $A_n$ is scattered away from the local origin $r = 0$, where the scatterer is centered, with amplitude $A_n C_n$. The coefficient $C_n$ acts as an amplification factor and is dimensionless. Further, each scattered mode has a $\mathrm{Log}\, r$ ($n = 0$) or $r^{-n}$ ($n \geq 1$) square-root singularity as $r$ approaches zero.

Observe that the origin $r = 0$ is inside the scatterer and the scattered field is defined outside the scatterer. Thus, there is no singularity in the region where (2.3) holds. The coefficients $C_n$ depend on frequency, as well as on the properties of the scatterer and those of the solid outside.

In the next section, we shall reduce the geometry of the scatterer to a line. To this line, we associate the scattered displacement (2.3), where the coefficients $C_n$ are evaluated by imposing appropriate continuity conditions on the boundary of an actual-size cylinder. Thus, the actual cylinder will be replaced with a new object; the geometries are different, but the scattering properties are identical. This geometrical simplification, however, causes the singularities of (2.3) to be no longer embedded inside the scatterer.

## 3. Coherent motion

The coherent motion in the solid of Fig. 1 is the sum of the incident motion and of the wave motions scattered by all the scatterers (Waterman & Truell 1961). In terms of the displacement, one has (Angel & Koba 1998)

$$u_2(y_3) = u_2^i(y_3) + n_0 \int_{-h}^{h} \int_{-\infty}^{\infty} u_2(y_3, y_1; x_3, x_1) dx_3 dx_1 . \qquad (3.1)$$

The second term on the right-hand side of (3.1) represents the global contribution of all scatterers located at all possible positions $(x_3, x_1)$ in the layer of thickness $2h$, under the assumption that the distribution is random and uniform. The dimension of the number density $n_0$ in (3.1) is an inverse area. The coordinate $y_1$ is integrated out on the right-hand side of (3.1) because of the translational invariance in the $y_1$ direction.

The term $u_2(y_3, y_1; x_3, x_1)$ in (3.1) is an average displacement corresponding to the wave motion scattered by the line at $(x_3, x_1)$. The average on disorder is taken over all configurations keeping the line at $(x_3, x_1)$ fixed. Thus, it is a partial average.

The exciting wave motion on a fixed scatterer is caused by scattering processes of all orders (single, double, triple…) (Ishimaru 1997) inside the layer. Keeping the scatterer fixed and averaging the displacements corresponding to those scattering processes, one defines the average exciting displacement $u_2^E$. Assuming that this displacement is a *bounded* solution of the two-dimensional Helmholtz equation in the solid with wavenumber $k$, one finds that the most general expression of $u_2^E$ corresponding to a scatterer fixed at $\boldsymbol{x}$ is

$$u_2^E(y_3, y_1; x_3, x_1) = u_0 \sum_{n=0}^{+\infty} \varepsilon_n E_n(x_3) J_n(k|\boldsymbol{y} - \boldsymbol{x}|) \cos(n\theta), \qquad (3.2)$$

where the angle $\theta$ is defined by $\cos\theta = (y_3 - x_3)/|\boldsymbol{y} - \boldsymbol{x}|$ and the distance $|\boldsymbol{y} - \boldsymbol{x}|$ is defined as in (2.3). The coefficients $E_n$ in (3.2) depend only on $x_3$, owing to the translational invariance in the $y_1$ direction.

Next we recall the scattering rule (2.3) and (2.5). Thus, we infer from (3.2) that the average scattered displacement discussed above is given by

$$u_2(y_3, y_1; x_3, x_1) = u_0 \sum_{n=0}^{+\infty} \varepsilon_n C_n E_n(x_3) H_n^{(1)}(k|\mathbf{y} - \mathbf{x}|)\cos(n\varphi). \quad (3.3)$$

Note that the singularity of $H_n^{(1)}$ at $\mathbf{y} = \mathbf{x}$ is not integrable over the two-dimensional domain $|x_1| < \infty$ and $|x_3| \leq h$ when $\mathbf{y}$ is inside the domain and when $n \geq 2$. Thus, the double integral of (3.1), with $u_2$ given by (3.3), is evaluated as in Waterman & Truell (1961), page 526. Namely, an entire infinite slab $|y_3 - x_3| < \varepsilon$ is omitted from the range of integration, and then the limit is taken as $\varepsilon$ approaches zero.

Without loss in generality in the present context, one can set $y_1 = 0$ on the right-hand side of (3.3). Substituting (3.3) into (3.1), one finds that the coherent displacement in the solid has the form

$$u_2(y_3) = u_2^i(y_3) + n_0 u_0 \sum_{n=0}^{+\infty} \varepsilon_n C_n \int_{-h}^{h} E_n(x_3) j_n(y_3 - x_3) dx_3. \quad (3.4)$$

The quantities $j_n$ on the right-hand side of (3.4) are defined by

$$j_n(y_3 - x_3) = \int_{-\infty}^{+\infty} H_n^{(1)}(k|\mathbf{y} - \mathbf{x}|)\cos(n\varphi) dx_1, \quad y_3 \neq x_3. \quad (3.5)$$

The integral (3.5) is evaluated over the entire real line. It can be evaluated by mathematical induction, which yields the results (Aguiar & Angel 2000; Urick & Ament 1949)

$$j_{2n}(y_3 - x_3) = (-1)^n \frac{2}{k} e^{ik|x_3 - y_3|}, \tag{3.6}$$

$$j_{2n+1}(y_3 - x_3) = i(-1)^n \frac{2}{k} \mathrm{sgn}(x_3 - y_3) e^{ik|x_3 - y_3|}, \tag{3.7}$$

where $n = 0, 1, 2 \ldots$ and sgn denotes the sign function. It follows from (3.4)-(3.7) that the coherent displacement in the solid is given by

$$u_2(y_3) = u_2^i(y_3) + in_0 u_0 \int_{-h}^{h} s_1(x)\mathrm{sgn}(x - y_3) e^{ik|x - y_3|} dx + n_0 u_0 \int_{-h}^{h} s_2(x) e^{ik|x - y_3|} dx. \tag{3.8}$$

The functions $s_1(x)$ and $s_2(x)$ in (3.8) are defined by

$$s_1(x) = \frac{4}{k} \sum_{j=0}^{+\infty} (-1)^j C_{2j+1} E_{2j+1}(x),$$

$$s_2(x) = \frac{2}{k} \sum_{j=0}^{+\infty} (-1)^j e_{2j} C_{2j} E_{2j}(x). \tag{3.9}$$

It follows from (3.8) that the reflected coherent motion is a plane wave of wavenumber $k$ and amplitude $u_0 R$ and the transmitted coherent motion is a plane wave of wavenumber $k$ and amplitude $u_0 T$, as in (2.2). Expressions for the coefficients $R$ and $T$ are obtained by setting, respectively, $y_3 < -h$ and $y_3 > h$ on the right-hand side of (3.8). We find that

$$R = n_0 \int_{-h}^{h} (s_2(x) + is_1(x)) e^{ikx} dx, \tag{3.10}$$

$$T = 1 + n_0 \int_{-h}^{h} (s_2(x) - is_1(x)) e^{-ikx} dx. \tag{3.11}$$

Inside the layer, where $-h < y_3 < h$, one finds from (3.8) that the coherent $u_2$ displacement looks rather complicated; it is the sum of one term with an $\exp(iky_3)$ factor and another term with an $\exp(-iky_3)$ factor. Both terms have "amplitudes" that depend on $y_3$. To clarify the situation, we need to determine the functions $E_n(x)$, which are still unknown at this point.

## 4. Scattered motion

For $-h < y_3 < h$, each of the two integrals in (3.8) is written as a sum of two integrals over the intervals $(-h, y_3)$ and $(y_3, h)$, respectively. Then, we write $y_3 - x = y_3 - z_3 + z_3 - x$, where $z_3$ is the coordinate along the $y_3$ direction of an arbitrary fixed point $z$ inside the layer. On each integration interval, we use the expansion

$$e^{ik(y_3 - z_3)} = \sum_{n=0}^{+\infty} e_n i^n J_n(k|y-z|)\cos(n\alpha), \qquad (4.1)$$

where the angle $\alpha$ is defined by $\cos\alpha = (y_3 - z_3)/|y - z|$. After some rather lengthy but straightforward calculations, one finds that (3.8), for $-h < y_3 < h$, can be written in the equivalent form

$$u_2(y_3) = u_2^i(y_3) + n_0 u_0 \sum_{n=0}^{+\infty} e_n G_n(y_3, z_3) J_n(k|y-z|)\cos(n\alpha). \qquad (4.2)$$

In this equation, the term $G_n$ can be written as the sum of a term that depends only on $z_3$ and another term that depends on both $z_3$ and $y_3$. Thus, one has

$$G_n(y_3, z_3) = F_n(z_3) + D_n(y_3, z_3), \qquad (4.3)$$

with $|y_3| < h$ and $|z_3| < h$. For even and odd values of the subscript, one finds that

$$F_{2n}(z_3) = (-1)^n F_0(z_3), \quad F_{2n+1}(z_3) = (-1)^n F_1(z_3), \tag{4.4}$$

$$D_{2n}(y_3, z_3) = (-1)^n D_0(y_3, z_3), \quad D_{2n+1}(y_3, z_3) = (-1)^n D_1(y_3, z_3), \tag{4.5}$$

$$F_0(z_3) = i\int_{-h}^{h} s_1(x)\,\text{sgn}(x-z_3)e^{ik|x-z_3|}dx + \int_{-h}^{h} s_2(x)e^{ik|x-z_3|}dx, \tag{4.6}$$

$$F_1(z_3) = \int_{-h}^{h} s_1(x)e^{ik|x-z_3|}dx - i\int_{-h}^{h} s_2(x)\,\text{sgn}(x-z_3)e^{ik|x-z_3|}dx, \tag{4.7}$$

$$D_0(y_3, z_3) = -2i\int_{z_3}^{y_3} s_1(x)\cos(k(x-z_3))dx - 2i\int_{z_3}^{y_3} s_2(x)\sin(k(x-z_3))dx, \tag{4.8}$$

$$D_1(y_3, z_3) = -2i\int_{z_3}^{y_3} s_1(x)\sin(k(x-z_3))dx + 2i\int_{z_3}^{y_3} s_2(x)\cos(k(x-z_3))dx. \tag{4.9}$$

The series expansion with the coefficients $G_n$ in (4.2) can be decomposed according to (4.3) into two series expansions. The series with the coefficients $F_n$ is a solution of the two-dimensional Helmholtz equation in the solid with wavenumber $k$. The series with the coefficients $D_n$ is not in general a solution of that equation, because $D_n$ depends on $y_3$. Observe that the expansion (4.2) holds also in $y_3 > h$ and $y_3 < -h$. In these intervals, the coefficients $G_n$ are independent of $y_3$. Exact expressions can be obtained readily from (3.8) and (4.1).

**5. Exciting motion**

The exciting wave motion on a fixed scatterer at $z$ is the sum of the incident wave and all the motions scattered by the other scatterers except that at $z$ (Ishimaru 1997). Averaging on disorder, keeping the scatterer at $z$ fixed, one finds that

$$u_2^E(y_3, y_1; z_3, z_1) = u_2^i(y_3) + n_0 \int_{-h}^{h} \int_{-\infty}^{+\infty} u_2(y_3, y_1; x_3, x_1; z_3, z_1) dx_3 dx_1. \tag{5.1}$$

On the right-hand side of (5.1), the integrand represents the average displacement scattered by the line at $(x_3, x_1)$ when the line at $(z_3, z_1)$ is kept fixed. The integral of this displacement over all positions $(x_3, x_1)$ in the layer of Fig. 1 represents the total contribution of all scattered motions knowing that the line at $(z_3, z_1)$ is fixed. The point at which the wave motion is evaluated is $(y_3, y_1)$. The average exciting displacement on the left-hand side of (5.1) is given by (3.2), where $(x_3, x_1)$ must be replaced by $(z_3, z_1)$.

The displacements $u_2^E$ and $u_2^i$ are bounded solutions of the two-dimensional Helmholtz equation in the solid with wavenumber $k$, where the derivatives in that equation are taken with respect to the $y$ coordinates. It follows that the integral in (5.1) is a bounded solution of the two-dimensional Helmholtz equation in the solid with wavenumber $k$. Thus, we write (5.1) in the equivalent form

$$u_2^E(y_3, y_1; z_3, z_1) = u_2^i(y_3) + n_0 u_0 \sum_{n=0}^{+\infty} e_n M_n(z_3) J_n(k|y-z|)\cos(n\alpha). \tag{5.2}$$

The coefficients $M_n$ in (5.2) depend only on $z_3$, owing to the translational invariance in the $y_1$ direction.

In (5.2), we expand $u_2^i$ by using the series (4.1), and the average exciting displacement $u_2^E$ by using (3.2), where $(x, \theta)$ are replaced with $(z, \alpha)$. All the terms in (5.2) are now expanded on the basis of functions $J_n(k|y-z|)\cos(n\alpha)$. Equating the coefficients on either side of (5.2), one finds that

$$E_n(z_3) = i^n e^{ikz_3} + n_0 M_n(z_3). \qquad (5.3)$$

In equation (5.3), the quantities $E_n(z_3)$ represent the average exciting displacement on a fixed scatterer at $z$ (averaging on the disorder of all the other scatterers). The quantities $n_0 M_n(z_3)$ represent a global scattered displacement, where the displacement scattered by a scatterer at $x$ is averaged on the disorder of all the other scatterers except those at $z$ and $x$ and, subsequently, all the positions $x$ in the layer are taken into account by the integration in (5.1). At this stage, neither $E_n$ nor $M_n$ are known.

## 6. Global closure assumption

Equation (5.1) expresses the exciting displacement on a fixed $z$ in terms of the scattered displacement corresponding to two fixed $x$ and $z$. Similarly, one can write an equation for the exciting displacement corresponding to two fixed $z$ and $u$ in terms of the scattered displacement keeping three scatterers fixed at $x$, $z$ and $u$. Repeating this process, one can construct a sequence of equations such that an additional scatterer is fixed at each step. Eventually, all the scatterers are fixed. This approach would in principle allow us to determine the quantities $E_n$ exactly but, unfortunately, it is too complicated and cannot be implemented in practice (Waterman & Truell 1961).

As an alternative to the approach described above, we propose a closure assumption that will allow us to determine the quantities $E_n$. The most simple assumption corresponds to the situation of simple scattering, where the scatterers are sufficiently distant from each other and the interactions between them can be neglected. In this case, it is assumed that the exciting displacement on each scatterer is equal to the incident displacement. The coefficients $E_n$ corresponding to simple scattering are obtained explicitly from (5.3) by setting the second term on the right-hand side of that equation equal to zero. Substituting these expressions for $E_n$ into (3.8), one finds the corresponding coherent displacement in the solid and, substituting into (3.10) and (3.11), one finds the reflection and transmission coefficients.

When interactions between scatterers cannot be neglected, one must take into account multiple scattering effects. These effects are represented by the quantities $M_n$ in (5.3). The $M_n$ can be viewed as a representation of the integral on the right-hand side of (5.1). Likewise, the $G_n$ in (4.2) can be viewed as a representation of the integral on the right-hand side of (3.1).

Comparing the integral in (3.1) with that in (5.1), we recall that the former is *not* a solution of the Helmholtz equation with wavenumber $k$ in the interval $|y_3| < h$, and the latter is indeed a solution of the Helmholtz equation with wavenumber $k$ for all $y_3$. Both integrals represent the same global scattered displacement averaged on disorder, except that in (5.1) one of the scatterers is fixed at $z$ and in (3.1) no scatterer is fixed.

In view of the preceding discussion, we assume that $M_n$ is equal to the part of $G_n$ that is a solution of the two-dimensional Helmholtz equation in the solid. Thus, referring to (4.3), we write

$$M_n(z_3) = F_n(z_3), \tag{6.1}$$

for $n = 0, 1, 2, \ldots$. The part of $G_n$ that is left out in (6.1) is $D_n$ which, we observe from (4.5), (4.8) and (4.9), approaches zero as $y_3$ approaches $z_3$. Equation (6.1) is our Global Closure Assumption.

To clarify the physical meaning of the assumption (6.1), we substitute (6.1) into (5.2), and then we recall (4.2) and (4.3). Then, we find that

$$u_2^E(y_3, y_1; z_3, z_1) = u_2(y_3) - n_0 u_0 \sum_{n=0}^{+\infty} \varepsilon_n D_n(y_3, z_3) J_n(k|y-z|)\cos n\alpha, \qquad (6.2)$$

where the functions $D_n$ are defined by (4.5), (4.8), and (4.9). Equation (6.2) can be viewed as an equivalent statement of the assumption (6.1); it provides an expression for the exciting displacement $u_2^E$ on a scatterer fixed at $z$ in terms of the coherent displacement $u_2$.

The expression (6.2) is defined at all points $y$ in the neighbourhood of $z$ and away from that neighbourhood. We see that $u_2^E$ approaches $u_2$ when $y$ approaches $z$, since the coefficients $D_n$ vanish when $y_3 = z_3$.

Substituting (6.1) into (5.3), and using (4.4), one finds that

$$E_{2n}(z_3) = (-1)^n E_0(z_3) \text{ and } E_{2n+1}(z_3) = (-1)^n E_1(z_3), \qquad (6.3)$$

with

$$E_0(z_3) = e^{ikz_3} + n_0 F_0(z_3), \qquad (6.4)$$

and

$$E_1(z_3) = i\, e^{ikz_3} + n_0 F_1(z_3). \qquad (6.5)$$

Equation (6.3) allows us to write the functions $s_1(x)$ and $s_2(x)$ in (3.9) only in terms of $E_0$ and $E_1$. We find that

$$s_1(x) = s_1 E_1(x) \text{ and } s_2(x) = s_2 E_0(x), \tag{6.6}$$

with

$$s_1 = \frac{4}{k} \sum_{j=0}^{+\infty} C_{2j+1}, \quad s_2 = \frac{2}{k} \sum_{j=0}^{+\infty} e_{2j} C_{2j}. \tag{6.7}$$

Substituting (6.6) into (4.6) and (4.7), one can write $F_0$ and $F_1$ in terms of $E_0$ and $E_1$. It follows that (6.4) and (6.5) represent a system of two coupled integral equations for the unknown functions $E_0$ and $E_1$.

## 7. Effective wavenumber

To simplify the notation, we use the variable $x$ in place of $z_3$. Then, with $|x| < h$, the system of equations for $E_0$ and $E_1$ has the form

$$E_0(x) = e^{ikx} + i n_0 s_1 \int_{-h}^{h} E_1(x') \operatorname{sgn}(x - x') e^{ik|x-x'|} dx' + n_0 s_2 \int_{-h}^{h} E_0(x') e^{ik|x-x'|} dx', \tag{7.1}$$

$$E_1(x) = i e^{ikx} + n_0 s_1 \int_{-h}^{h} E_1(x') e^{ik|x-x'|} dx' - i n_0 s_2 \int_{-h}^{h} E_0(x') \operatorname{sgn}(x - x') e^{ik|x-x'|} dx'. \tag{7.2}$$

We differentiate (7.1) and (7.2) two times with respect to $x$, keeping the variable $x$ in the interval $(-h, h)$. We use the properties that the derivatives of $\operatorname{sgn}(x)$ and $e^{|x|}$, respectively, are $2\delta(x)$ and $\operatorname{sgn}(x) e^{|x|}$, where $\delta(x)$ denotes the Dirac delta function. Thus, we find that

$$\frac{d^2}{dx^2}E_0(x) = -K^2 E_0(x) \quad \text{and} \quad \frac{d^2}{dx^2}E_1(x) = -K^2 E_1(x). \tag{7.3}$$

The coefficient $K$ is independent of $x$ and is given by

$$K^2 = (k - 2in_0 s_1)(k - 2in_0 s_2). \tag{7.4}$$

The general solutions of the second-order ordinary differential equations (7.3) are

$$E_0(x) = V_0 e^{iKx} + W_0 e^{-iKx}, \tag{7.5}$$

$$E_1(x) = V_1 e^{iKx} + W_1 e^{-iKx}, \tag{7.6}$$

where the coefficients $V_0$, $W_0$, $V_1$, and $W_1$ are independent of $x$. We now infer from (3.8), (6.6), (7.1), and (7.5) that the coherent displacement in the interval $|y_3| < h$ is given by

$$u_2(y_3) = u_0 E_0(y_3) = u_0 A_+ e^{iKy_3} + u_0 A_- e^{-iKy_3}, \tag{7.7}$$

where $A_+ = V_0$ and $A_- = W_0$. Equation (7.7) shows that, inside the layer where the scatterers are distributed, there is one wave traveling in the forward direction and a second wave traveling in the backward direction. Both waves are governed by $K$, which is the effective wavenumber. In general, the wavenumber $K$ is complex-valued and depends on frequency. It follows that the two waves inside the layer are dispersive and undergo attenuation as they propagate.

It is customary to write the effective wavenumber $K$ of (7.4) in terms of the far-field scattered displacement corresponding to a single scatterer subjected to the incident wave of (2.2) (Waterman & Truell 1961). To arrive at the usual formula for $K$, we recall that the cylindrical Hankel function is given in the far-field as $r$ tends to infinity by the expression

$$H_n^{(1)}(kr) = (-i)^n \sqrt{\frac{2}{\pi kr}} e^{i\left(kr - \frac{\pi}{4}\right)}. \tag{7.8}$$

Substituting (2.4) and (7.8) into (2.3), one finds that

$$u_2(r,\theta) = u_0 \sqrt{\frac{2\pi}{kr}} e^{i\left(kr + \frac{\pi}{4}\right)} f(\theta). \tag{7.9}$$

When the scatterer is located at $\mathbf{x} = \mathbf{0}$, the angular shape function $f$ in (7.9) is given by

$$f(\theta) = \frac{1}{i\pi} \sum_{n=0}^{+\infty} \varepsilon_n C_n \cos(n\theta). \tag{7.10}$$

We can see from Fig. 1 that, relative to the incident wave, the forward direction corresponds to $\theta = 0$ and the backward direction to $\theta = \pi$. It follows that $f(0)$ and $f(\pi)$, respectively, are measures in the forward and backward directions of the displacement scattered by a single scatterer. We infer from (6.7) and (7.10) that

$$s_1 = \frac{i\pi}{k}(f(0) - f(\pi)) \text{ and } s_2 = \frac{i\pi}{k}(f(0) + f(\pi)). \tag{7.11}$$

Equation (7.11) shows that the factors $s_1$ and $s_2$ depend only on the far-field amplitudes scattered by a single scatterer in the forward and backward directions, apart from the wavenumber $k$. Substituting (7.11) into (7.4), and rearranging terms, one finds the well-known formula (Waterman & Truell 1961)

$$(7.12) \qquad K^2 = k^2 \left(1 + n_0 \frac{2\pi}{k^2} f(0)\right)^2 - k^2 \left(n_0 \frac{2\pi}{k^2} f(\pi)\right)^2.$$

## 8. Reflection and transmission

We determine the coefficients $V_0$, $W_0$, $V_1$, and $W_1$ in closed form by substituting (7.5) and (7.6) into (7.1) and (7.2), as in Angel & Aristégui (2005). Next, we substitute these coefficients, together with (6.6), (7.5), (7.6), into (3.10) and (3.11). Integrating the exponential functions in closed form, we find that the reflection and transmission coefficients are given by

$$R = \frac{Q\, e^{-2ikh}}{1 - Q^2\, e^{4iKh}} \left(1 - e^{4iKh}\right), \tag{8.1}$$

$$T = \frac{1 - Q^2}{1 - Q^2\, e^{4iKh}} e^{2i(K-k)h}, \tag{8.2}$$

where

$$Q = \frac{1 - \mathcal{Q}}{1 + \mathcal{Q}} \tag{8.3}$$

and

$$\mathcal{Q} = \frac{k - 2in_0 s_2}{K} = \frac{K}{k - 2in_0 s_1}. \tag{8.4}$$

Outside the screen, the reflected and transmitted waves have amplitudes $u_0 R$ and $u_0 T$, respectively, where $R$ and $T$ are given by (8.1) and (8.2). Inside the screen, where $|y_3| < h$, the forward and backward waves have amplitudes $u_0 A_+$ and $u_0 A_-$, respectively, where $A_+$ and $A_-$ are given by

$$V_0 = A_+ = \frac{1 + Q}{1 - Q^2\, e^{4iKh}} e^{i(K-k)h}, \tag{8.5}$$

$$W_0 = A_- = -\frac{Q(1+Q)}{1-Q^2 e^{4iKh}} e^{i(3K-k)h}. \tag{8.6}$$

The factor $Q$ in (8.1), (8.2), (8.5) and (8.6) is defined by (8.3) and (8.4), together with (6.7) or (7.11).

The formulae (8.1), (8.2), (8.5) and (8.6) for the reflected, transmitted, forward, and backward waves, respectively, apply also to a homogeneous elastic plate. To see this, replace in Fig. 1 the screen of scatterers with a homogeneous elastic plate of thickness $2h$ and mass density $r_1$. In this plate, let $k_1 = w/c_1$ be the wavenumber, where $c_1$ is the speed of transverse waves. Then, one finds, by imposing continuity of stress and displacement on the boundaries $y_3 = \pm h$, that the amplitudes of the four waves generated by the incident wave (2.2) are still given by (8.1), (8.2), (8.5) and (8.6), provided that $K$ be replaced with $k_1$ and $Q$ with

$$Q = \frac{r_0 c_0 - r_1 c_1}{r_0 c_0 + r_1 c_1}. \tag{8.7}$$

The factor $Q$ is a ratio of acoustic impedances. This shows that the system of waves propagating in the presence of scatterers in Fig. 1 is formally identical to that propagating outside and inside a homogeneous elastic plate. The effective wavenumber $K$ is replaced with $k_1$; the factor $Q$ of (8.3) and (8.4) is replaced with (8.7).

## 9. Comparison between SH and P solutions

We compare the solution obtained in this paper – for SH waves with particle displacements along the $y_2$ direction – with that obtained in an earlier work (Angel & Aristégui 2005) – for P waves with particle displacements along the $y_3$ direction. In both cases, the geometry is that of Fig. 1 and the scatterers have cylindrical geometry with axes parallel to the $y_2$ direction.

In both cases, after making a closure assumption, the formulation of the scattering problem reduces to a pair of coupled integral equations for two unknown functions $E_0$ and $E_1$. For the purpose of this discussion, we call $E_0^S$ and $E_1^S$ the solutions of (7.1) and (7.2); we call $E_0^P$ and $E_1^P$ the solutions of (56) and (57) in the P-wave work (Angel & Aristégui 2005). Likewise, the factors $s_1$ and $s_2$ are labeled, respectively, with an $S$ or a $P$ superscript. Then, if $E_0^S$ and $E_1^S$ are solutions of (7.1) and (7.2), one infers that

$$E_0^P = \frac{1}{ik} E_0^S, \quad E_1^P = \frac{1}{ik} E_1^S, \tag{9.1}$$

are solutions of (56) and (57) in the P-wave work (Angel & Aristégui 2005). The solution of the system (7.1) and (7.2) is

$$E_0^S(x) = V_0^S e^{iKx} + W_0^S e^{-iKx}, \tag{9.2}$$

$$E_1^S(x) = V_1^S e^{iKx} + W_1^S e^{-iKx}, \tag{9.3}$$

where the wavenumber $K$ is given by (7.4) and no superscript is attached to $K$ for simplicity. The coefficients in (9.2) and (9.3) are given by

$$V_0^S = (1 + Q^S) e^{i(K-k)h} / D^S, \tag{9.4}$$

$$W_0^S = -Q^S (1 + Q^S) e^{i(3K-k)h} / D^S, \tag{9.5}$$

$$V_1^S = i Q^S V_0^S, \tag{9.6}$$

$$W_1^S = -i\, Q^S\, W_0^S, \tag{9.7}$$

where

$$D^S = 1 - \left(\mathcal{Q}^S\right)^2 e^{4iKh}, \tag{9.8}$$

$$\mathcal{Q}^S = \frac{1 - Q^S}{1 + Q^S}, \tag{9.9}$$

$$Q^S = \frac{k - 2in_0 s_2}{K} = \frac{K}{k - 2in_0 s_1}. \tag{9.10}$$

In equations (9.4)-(9.10), the superscript $S$ has been omitted on $K$, $s_1$ and $s_2$. It follows from (9.1) that the solutions (9.2) and (9.3) – with the superscript $P$ replacing $S$ – can be used to determine $E_0^P$ and $E_1^P$. Then, one infers from (9.1) and (9.4)-(9.10) that

$$ikV_0^P = \left(1 + \mathcal{Q}\right) e^{i(K-k)h} / D^P, \tag{9.11}$$

$$ikW_0^P = -\mathcal{Q}\left(1 + \mathcal{Q}\right) e^{i(3K-k)h} / D^P, \tag{9.12}$$

$$V_1^P = i\, Q^P\, V_0^P, \tag{9.13}$$

$$W_1^P = -i\, Q^P\, W_0^P, \tag{9.14}$$

where

$$D^P = 1 - \mathcal{Q}^2 e^{4iKh}, \tag{9.15}$$

$$\mathcal{Q} = \frac{1 - Q^P}{1 + Q^P}, \tag{9.16}$$

$$Q^P = \frac{k - 2in_0 s_2}{K} = \frac{K}{k - 2in_0 s_1}. \tag{9.17}$$

In equations (9.11)-(9.17), the superscript $P$ has been omitted on $K$, $s_1$ and $s_2$. Next, as in equation (76) of the P-wave work (Angel & Aristégui 2005), we define $Z$ and $Q^P$ by

$$Z = \frac{1}{Q^P},\ Q^P = \frac{1-Z}{1+Z}.\tag{9.18}$$

Then, it follows from (9.16) and (9.18) that

$$\overset{\backslash}{Q} = -Q^P.\tag{9.19}$$

Using (9.18) and (9.19), one infers from (9.11)-(9.15) that

$$ikV_0^P = (1-Q^P)e^{i(K-k)h}/D^P,\tag{9.20}$$

$$ikW_0^P = Q^P(1-Q^P)e^{i(3K-k)h}/D^P,\tag{9.21}$$

$$kV_1^P = (1+Q^P)e^{i(K-k)h}/D^P,\tag{9.22}$$

$$kW_1^P = -Q^P(1+Q^P)e^{i(3K-k)h}/D^P,\tag{9.23}$$

where

$$D^P = 1-(Q^P)^2 e^{4iKh}.\tag{9.24}$$

Next, recall from (7.7) that the SH displacement inside the layer is proportional to $E_0$. Thus, it follows from (7.7) and (9.2) that for SH waves

$$A_+ = V_0^S,\ A_- = W_0^S.\tag{9.25}$$

Equation (9.25), together with (9.4), (9.5) and (9.8), is in agreement with (8.5) and (8.6).

In the case of P waves, the displacement inside the layer is proportional to $E_1$, as can be seen from (62) in the P-wave work (Angel & Aristégui 2005). Thus, it follows from (62) and (77) in Angel & Aristégui (2005), and from (9.3) with the superscript $S$ replaced with $P$, that for P waves

$$A_+ = kV_1^P, \quad A_- = -kW_1^P. \tag{9.26}$$

Equation (9.26), together with (9.22), (9.23) and (9.24), is in agreement with (78) and (79) in Angel & Aristégui (2005).

We observe that the sign conventions for the amplitudes of the backward traveling waves inside the layer are opposite, respectively, for the cases of SH and P waves. This can be seen from (7.7) in this paper and from (77) in Angel & Aristégui (2005).

From (3.10), (3.11), and (6.6), the reflection and transmission coefficients for SH waves are given by

$$R = n_0 \int_{-h}^{h} \left( s_2 E_0^S(x) + is_1 E_1^S(x) \right) e^{ikx} dx, \tag{9.27}$$

$$T = 1 + n_0 \int_{-h}^{h} \left( s_2 E_0^S(x) - is_1 E_1^S(x) \right) e^{-ikx} dx. \tag{9.28}$$

In (9.27) and (9.28), the superscript $S$ has been omitted on $R$, $T$, $s_1$ and $s_2$.

For P waves, we find expressions for $R$ and $T$ from equations (24), (25), and (53) in Angel & Aristégui (2005). The results are

$$R = ikn_0 \int_{-h}^{h} (s_2 E_0^P(x) + is_1 E_1^P(x)) e^{ikx} dx, \quad (9.29)$$

$$T = 1 + ikn_0 \int_{-h}^{h} (s_2 E_0^P(x) - is_1 E_1^P(x)) e^{-ikx} dx. \quad (9.30)$$

In (9.29) and (9.30), the superscript $P$ has been omitted on $R$, $T$, $s_1$ and $s_2$.

Substituting (9.2)-(9.10) into (9.27) and (9.28), one finds expressions for $R$ and $T$ that are those of (8.1) and (8.2), with a superscript $S$ attached to the $Q$ factor.

For P waves, expressions for $R$ and $T$ are obtained by substituting (9.1) into (9.29) and (9.30), and using the preceding results for (9.27) and (9.28). Then, with the factor $\breve{Q}$ of (9.16), one finds that

$$R = \breve{Q} \, e^{-2ikh} (1 - e^{4iKh}) / D^P, \quad (9.31)$$

$$T = (1 - \breve{Q}^2) e^{2i(K-k)h} / D^P, \quad (9.32)$$

where the superscript $P$ is omitted on $R$, $T$, and $K$. Now, we recall from (9.19) that $Q^P = -\breve{Q}$. Then, in terms of $Q^P$, we find that (9.31) and (9.32) are identical to (74) and (75) in Angel & Aristégui (2005).

We observe that the sign conventions for the amplitudes of the reflected waves are opposite, respectively, for the cases of SH and P waves. This can be seen from (2.2) in this paper and from (2) in Angel & Aristégui (2005).

We have shown in this Section that the SH solution and the P solution are related. However, they cannot be inferred easily from each other. Indeed, the P problem requires that scalar

potentials – from which displacements follow by differentiation – be considered. For the SH problem, we work directly with the displacements, which are solutions of scalar Helmholtz equations.

The scattering properties of a single cylindrical scatterer are described by coefficients $C_n$. These $C_n$ serve to represent the scattered displacement in the SH problem and the scattered displacement potential in the P problem. The quantities $s_1$ and $s_2$ are series of the coefficients $C_n$, with the same formal expressions in either case – as can be seen from (6.7) in this paper and from (54) in Angel & Aristégui (2005), respectively.

The amplitudes $A_-$, $A_+$, $R$, and $T$ have the same forms, respectively, for SH and P waves – except for the sign changes concerning $A_-$ and $R$, which reflect only our choice of two different sign conventions for backward-traveling waves. This formal identity in the expressions of the two sets of amplitudes is obtained by introducing two distinct definitions for the factors $Q^S$ and $Q^P$. These two factors are *formally* opposite of each other, because $\mathrm{Q}^S$ and $Z$ are inverses of each other. Indeed, we recall from (9.9)-(9.10) and from (9.17)-(9.18), respectively, that

$$Q^S = \frac{1 - \mathrm{Q}^S}{1 + \mathrm{Q}^S}, \; \mathrm{Q}^S = \frac{k - 2\mathrm{i}n_0 s_2}{K}, \tag{9.33}$$

$$Q^P = \frac{1 - Z}{1 + Z}, \; Z = \frac{K}{k - 2\mathrm{i}n_0 s_2}. \tag{9.34}$$

The results are presented in terms of $Q^S$ for SH waves and of $Q^P$ for P waves because they become identical in form to those obtained for homogeneous plates. Indeed, we have shown that the formulae for $A_-$, $A_+$, $R$, and $T$ are still valid for homogeneous plates, provided that $K$ be replaced with $k_1$ (the wavenumber in the plate) and $Q$ be interpreted as a ratio of real-valued impedances. In either case ($Q^S$ or $Q^P$), one has

$$Q = \frac{r_0 c_0 - r_1 c_1}{r_0 c_0 + r_1 c_1}. \tag{9.35}$$

In equation (9.35), $r_0$ is the mass density of the material outside the layer and $r_1$ is the mass density of the homogeneous layer. For SH waves, we have two solids; for P waves, two fluids. The speeds $c_0$ and $c_1$, outside and inside the layer respectively, are either speeds of transverse waves or of longitudinal waves.

## 10. Cracks and cavities

We illustrate the results of Section 8 first when the scatterers are flat cracks and then when they are cylindrical cavities with circular cross-sections. We focus attention on the reflection and transmission coefficients $R$ and $T$ of equations (8.1) and (8.2). We shall compare the results with those obtained elsewhere by different approaches.

The cracks have width $2a$ and infinite length along the $y_2$ direction. Their faces lie in planes parallel to the $(y_1, y_2)$ plane. When cracks are represented as lines of discontinuity of negligible thickness, the solution of the one-crack problem cannot be expressed in closed form. It can be expressed in terms of a dislocation density, which is a solution of a singular integral equation (Angel 1988).

For a crack with a cross-section located along the segment $y_3 = 0$, $|y_1| < a$, in the $(y_3, y_1)$ plane, the dislocation density $b$ satisfies the two equations

$$\int_{-a}^{a} b(s) \left\{ \frac{1}{s - y_1} + S(s - y_1) \right\} ds = i\rho u_0 k, \quad |y_1| < a, \tag{10.1}$$

$$\int_{-a}^{a} b(s) ds = 0. \tag{10.2}$$

In equation (10.1), the function $S$ is given by

$$S(x) = \int_{0}^{\infty} \left( \frac{\beta}{x} - 1 \right) \sin(xx) dx, \tag{10.3}$$

where $\beta$ is defined by $\beta^2 = x^2 - k^2$ and $\operatorname{Im} \beta \leq 0$. In the solid containing this crack, the steady-state scattered displacement $u_2$ can be expressed in the form (Angel 1988)

$$u_2(y_1, y_3) = \frac{i}{2} \operatorname{sgn}(y_3) k y_3 \int_{-a}^{a} \frac{u_2(s, 0^+)}{r_0} H_1^{(1)}(k r_0) ds, \tag{10.4}$$

where $r_0$ is defined by $r_0^2 = (y_1 - s)^2 + y_3^2$. In (10.4), sgn denotes the sign function.

Let $(r, q)$ be the polar coordinates of an arbitrary point in Fig. 1, with the angle $q$ measured from the $y_3$ axis in the clockwise direction and the origin at the intersection of the $y_3$ and $y_1$ axes. Then, in the far-field, one infers from (10.4) and the asymptotic expression (7.8) of the Hankel function that

$$u_2(y_1, y_3) = u_2(r, q) = u_0 \sqrt{\frac{2\pi}{kr}} e^{i\left(kr + \frac{\pi}{4}\right)} f(q), \quad (r \gg a, 0 \leq q < 2\pi). \tag{10.5}$$

In equation (10.5), $f$ is the angular shape function of the crack. It is defined by

$$f(q) = \frac{\cot q}{2\rho u_0} \int_{-a}^{a} b(s) e^{-iks \sin q} ds.  \tag{10.6}$$

The crack-opening displacement $u_2(y_1, 0^+)$ in (10.4) and the dislocation density $b$ are related by

$$u_2(y_1, 0^+) = \int_{y_1}^{a} b(s) ds, \quad |y_1| < a,  \tag{10.7}$$

$$u_2(y_1, 0^+) = 0, \quad |y_1| \geq a.  \tag{10.8}$$

Since $u_2(y_1, 0^+)$ is an even function of $y_1$, it follows from (10.7) that $b(y_1)$ is an odd function of $y_1$ in the interval $|y_1| < a$. From this observation and from (10.6), we infer that $f(q) = f(-q)$ and $f(\pi - q) = -f(q)$. Taking the limit of (10.6) as $q$ approaches zero, and using (10.2), one finds that

$$f(0) = -f(\pi) = \frac{k}{2i\rho u_0} \int_{-a}^{a} s b(s) ds.  \tag{10.9}$$

For cracks distributed in the layer of Fig. 1, one finds from (7.4), (7.11), and (10.9) that the effective wavenumber $K$ is given by

$$K^2 = k\left(k - 2i\frac{n_0}{u_0} \int_{-a}^{a} s b(s) ds\right).  \tag{10.10}$$

From (10.9), together with (7.11), (8.3) and (8.4), one infers that

$$Q = \frac{K - k}{K + k},  \tag{10.11}$$

where $K$ is the complex root of (10.10) that lies in the first quadrant. Next, we determine the reflection and transmission coefficients $R$ and $T$ by substituting (10.10) and (10.11) into (8.1) and (8.2).

In Figs. 2 and 3, respectively, we show the modulus of the reflection coefficient and that of the transmission coefficient versus the dimensionless frequency $\omega = ka$. The cracks are distributed in a layer of thickness $h/a = 3$; their concentrations for the three curves shown on either figure are $e = n_0 a^2 = 0.01, 0.03, 0.05,$ respectively. Figures 2 and 3 are in very good agreement with Figs. 5 and 7 in Angel & Koba (1998).

In the approach of Angel and Koba (1998), boundary conditions for the average exciting field are written on each finite-size crack. The cracks are not replaced with line scatterers, as is done in this paper. Using our notations, the formula of Angel & Koba (1998) for the complex-valued wavenumber can be written in the form

$$K^2 = k^2 \left( 1 + 2i \frac{n_0}{k u_0} \int_{-a}^{a} sb(s) ds \right)^{1}. \tag{10.12}$$

We can see that (10.10) and (10.12) yield identical expressions to within first-order terms in $e$. Thus, the results of this paper and those of Angel & Koba (1998) are expected to coincide for low values of $e$. In fact, Figures 2 and 3 show that small differences can be seen for $e = 0.05$. For $e = 0.01$ or $0.03$, the curves of this paper and those of Angel & Koba (1998) nearly superimpose.

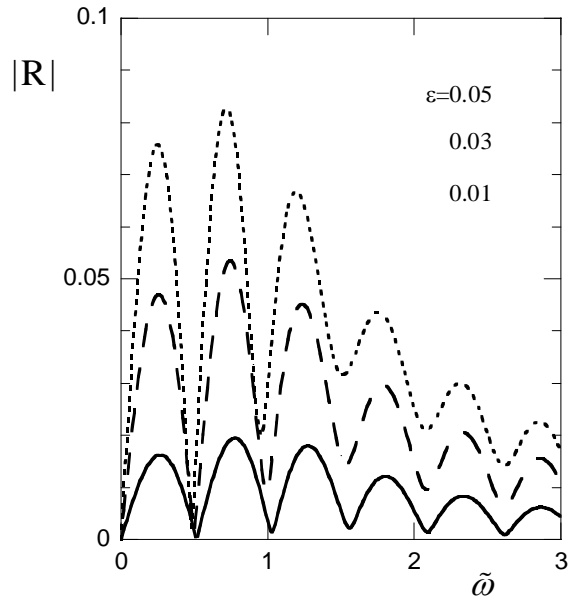

Figure 2 : Modulus of the reflection coefficient versus the dimensionless frequency for dimensionless thickness $h/a = 3$ and crack densities $e = 0.01$, $0.03$, $0.05$.

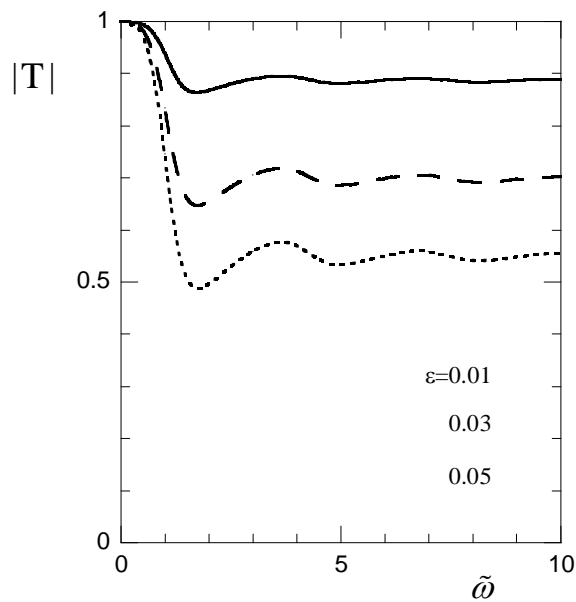

Figure 3 : Modulus of the transmission coefficient versus the dimensionless frequency for dimensionless thickness $h/a = 3$ and crack densities $e = 0.01$, $0.03$, $0.05$.

Figure 4 shows the modulus of the reflection coefficient for a distribution of cracks occupying a half-space. In this case, one finds that (2.2) and (8.1) yield

$$R = Q. \qquad (10.13)$$

For cylindrical cavities of radius $a$ and infinite length along the $y_2$ direction, we determine first the scattering coefficients $C_n$ corresponding to a single cavity. By using a Fourier-series representation and solving a boundary-value problem, where the shear stress vanishes on the boundary of the cavity, one finds that

$$C_n = -\frac{J_n'(\tilde{\omega})}{H_n^{(1)'}(\tilde{\omega})}, \quad n \geq 0. \qquad (10.14)$$

In equation (10.14), the prime superscript denotes differentiation.

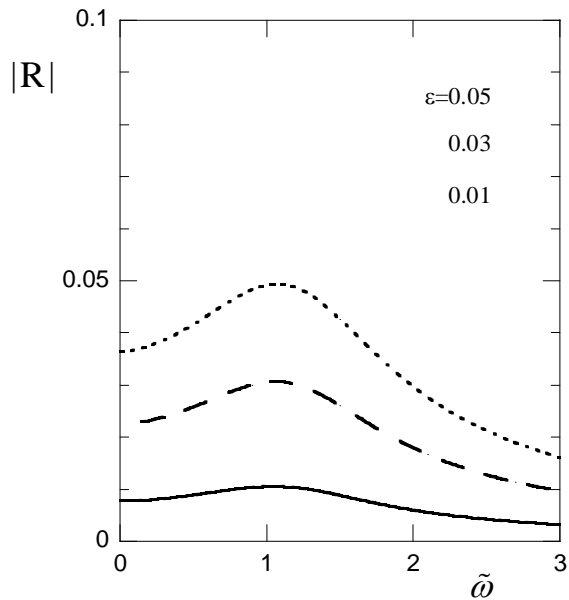

Figure 4 : Modulus of the reflection coefficient versus the dimensionless frequency for a half-space containing flat cracks. The crack density is $\varepsilon = 0.01, 0.03, 0.05$.

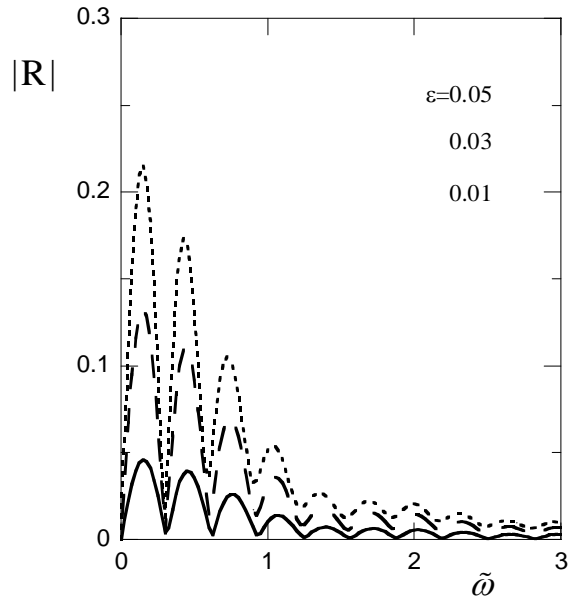

Figure 5 : Modulus of the reflection coefficient versus the dimensionless frequency for dimensionless thickness $h/a = 5$ and cavity densities $e = 0.01, 0.03, 0.05$.

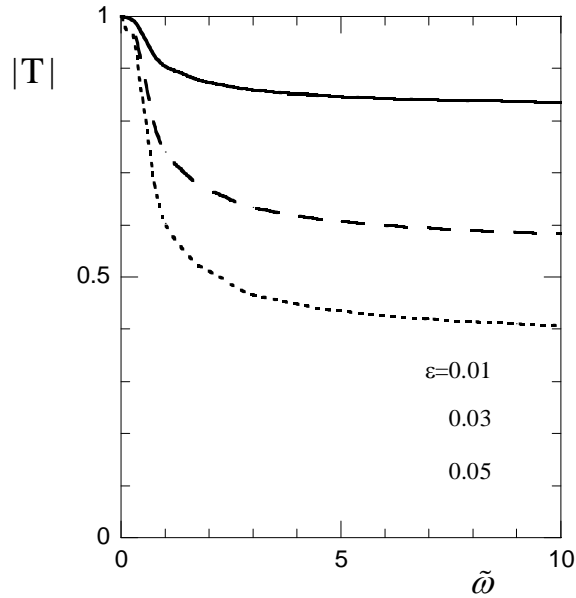

Figure 6 : Modulus of the transmission coefficient versus the dimensionless frequency for dimensionless thickness $h/a = 5$ and cavity densities $e = 0.01, 0.03, 0.05$.

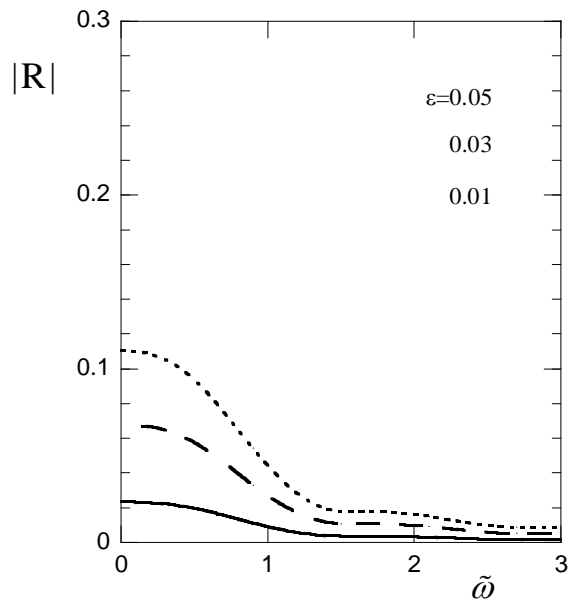

Figure 7 : Modulus of the reflection coefficient versus the dimensionless frequency for a half-space containing cylindrical cavities. The cavity density is $e = 0.01$, $0.03$, $0.05$.

With $K$, $s_1$, and $s_2$ determined from (10.14), (6.7), and (7.4), we use (8.1)-(8.4) to evaluate the reflection and transmission coefficients $R$ and $T$. In the series $s_1$ and $s_2$, we take the first fifty terms, which yields convergence of the respective sums in the frequency range $0 < \tilde{\omega} \leq 10$.

Figures 5 and 6 show, respectively, the modulus of $R$ and that of $T$ versus $\tilde{\omega}$. The cavities are distributed in a layer of thickness $h/a = 5$; their concentrations are $e = n_0 a^2 = 0.01$, $0.03$, and $0.05$. We find that these results are nearly identical to those of Aguiar & Angel (2000) for low cavity concentration.

In the approach of Aguiar and Angel (2000), boundary conditions for the average exciting field are written on each finite-size cavity. As a result, the reflection and transmission coefficients in Aguiar & Angel (2000) are represented by Fourier series and the precise numerical evaluation of those series is time-consuming. In this paper, the cavities are replaced with line scatterers by using the coefficients $C_n$ of (10.14). Figure 7 shows the modulus of the reflection coefficient for a distribution of cavities occupying a half-space. In this case, equation (10.13) holds.

## 11. Conclusion

For normal incidence, we have solved analytically the problem of scattering of an antiplane (SH) time-harmonic wave by a region of finite-thickness containing a distribution of cylindrical scatterers. The scatterers are identical. In each unit of area, in a plane perpendicular to the cylinder axes, the number of scatterers is equal to a constant $n_0$.

We have determined the effective wavenumber $K$, and the amplitudes of the four coherent SH waves inside and outside the scattering layer. These amplitudes, which are denoted by $R$, $T$, $A_+$, and $A_-$, are given in closed form in equations (8.1), (8.2), (8.5), and (8.6), respectively.

To arrive at these results, we have used the Twersky-Foldy (TF) and Waterman-Truell (WT) approaches. We find that the amplitudes $R$, $T$, $A_+$, and $A_-$ are completely determined – for a given solid, a given scatterer distribution, and a given frequency – when the backscattered amplitude $f(\pi)$ and the forward-scattered amplitude $f(0)$ of a single scatterer are known.

This is a fundamental result of TF and WT. It rests on a closure assumption, because the formulation of the multiple-scattering problem yields a system of an infinite number of equations that cannot be solved either analytically or numerically. Thus, in order to arrive at some useful analytical expressions, it is necessary to truncate – or close – the system of equations. The closure assumption of TF and WT makes use of translational addition theorems for cylindrical wavefunctions. In this paper, we have used a different approach, which is illustrated by equations (5.3) and (6.1), and which yields the same expression for the effective wavenumber as that of WT.

This paper shows that wave amplitudes inside and outside the layer can be obtained without imposing continuity conditions on the "boundaries" of the layer. In fact, the layer has no physical boundaries, since the solid is one and the same inside and outside. It would be erroneous to impose conditions on the "boundaries" – even though numerical results in particular situations could be nearly identical with those obtained in the context of the present work.

We have shown that our formulae for $R$ and $T$ can be easily applied to thin cracks and cylindrical cavities. In each case, we need only recall the solution of the single-scatterer problem. The results for cracks and cavities are in very good agreement with earlier results obtained by more labor-intensive methods.